\documentclass[conference]{IEEEtran}
\IEEEoverridecommandlockouts
\usepackage{amsmath,amssymb,amsfonts}
\usepackage{algorithmic}
\usepackage{algorithm2e}
\usepackage{graphicx}
\usepackage{textcomp}
\usepackage{xcolor}
\usepackage{comment}
\def\BibTeX{{\rm B\kern-.05em{\sc i\kern-.025em b}\kern-.08em
    T\kern-.1667em\lower.7ex\hbox{E}\kern-.125emX}}

\usepackage{biblatex}
\RestyleAlgo{ruled}
\begin{document}

\title{DHNet: A Distributed Network Architecture for Smart Home}

\author{\IEEEauthorblockN{1\textsuperscript{st} Chaoqi Zhou}
\IEEEauthorblockA{\textit{Southern University of} \\
\textit{Science and Technology} \\
ShenZhen, China \\
12133095@mail.sustech.edu.cn}

\and

\IEEEauthorblockN{2\textsuperscript{nd} Jingpu Duan}
\IEEEauthorblockA{\textit{Peng Cheng Laboratory} \\
ShenZhen, China \\
duanjp@pcl.ac.cn}

\and

\IEEEauthorblockN{3\textsuperscript{nd} YuPeng Xiao}
\IEEEauthorblockA{\textit{Peng Cheng Laboratory} \\
ShenZhen, China}

\and

\IEEEauthorblockN{4\textsuperscript{rd} Qing Li}
\IEEEauthorblockA{\textit{Peng Cheng Laboratory} \\
ShenZhen, China \\
liq@pcl.ac.cn}
\and
\IEEEauthorblockN{5\textsuperscript{th} Dingding Chen}
\IEEEauthorblockA{\textit{Huawei Technologies Co., Ltd} \\
ShenZhen, China \\
chendingding5@huawei.com}
\and
\IEEEauthorblockN{6\textsuperscript{th} Ruobin Zheng}
\IEEEauthorblockA{\textit{Huawei Technologies Co., Ltd} \\
ShenZhen, China \\
zhengruobin@huawei.com}

\and
\IEEEauthorblockN{7\textsuperscript{th} Shaoteng Liu}
\IEEEauthorblockA{\textit{Huawei Technologies Co., Ltd} \\
ShenZhen, China }

}

\maketitle

\begin{abstract}

Smart home are rapidly emerging as a focus area for future, aiming to provide ubiquitous intelligence services for daily household life, which will fundamentally changes our interactions with technology in living environment. The idea of the ubiquitous intelligent service involves the seamless integration of services into the background, ensuring that users can enjoy these services without being consciously aware of their presence. Undoubtedly, the communication latency of home networks is a critical factor in creating such pervasive intelligent experience. 
Traditional home networks usually adopt a centralized network topology because of their small scale, typically connecting just a few to dozens of terminals. In a centralized network topology, all terminal devices are connected to a core router either through wired or wireless, imposing all information to be routed and forwarded through the core router. However, in the age of smart homes,with the proliferation of connected devices and frequent data exchange between devices and the cloud, the core router can easily become overwhelmed by the influx of demands, resulting in network congestion and delays, ultimately impacting user experience negatively.
Therefore, decentralized networking solutions have become the best choice for smart home networking. However, there are several issues with the IP-based routing approach in decentralized networks: a) Each forwarding of an IP packet requires a lookup in the routing table. As the routing table grows, the lookup speed gradually decreases, leading to increased forwarding latency. b) During roaming, a new IP address needs to be reassigned, causing interruptions in IP network applications and requiring expensive means to restore the network. c) Maintaining routing tables and assigning IP addresses for each network device increases network management costs. To address these issues, we propose a new network protocol – the Path Vector Header (PVH) protocol. Compared to the IP network protocol, the new protocol, while fully compatible with the IP network protocol, reduces end-to-end latency in smart home networks by xx and improves average forwarding speed by xx times with only a small amount of bandwidth overhead. Additionally, since the PVH protocol uses a source routing algorithm, it completely eliminates the need for routing tables in the network, achieving a service-centric network routing approach.

\end{abstract}

\begin{IEEEkeywords}
networking, protocol, source routing, cluster
\end{IEEEkeywords}

\section{Introduction}

With the widespread popularity of smart homes\cite{perera2015emerging}, a large number of smart devices have entered ordinary households, bringing a ubiquitous convenient and intelligent experience to home life. Common smart home devices include terminal devices such as smartphones and tablets, household appliances such as refrigerators and washing machines, and household facilities such as sensors and lighting fixtures.
Smart home can mainly be divided into three parts, namely smart home system, smart items and smart devices. Smart home system mainly plays the role of induction, link and control, while smart single products give more intelligent functions to traditional home products. At present, the smart home market presents a growing trend. The volume growth indicates that the smart home industry has entered an inflection point and begin the next round of integration evolution. In the last three to five years, smart home has reached a relatively rapid development stage, and protocols and technical standards began to actively integrate with each other.

The work of smart home system relies on stable and reliable network environment\cite{choudhary2016internet, razzaque2015middleware}. The existing centralized home network relies too much on the central network device. With the increase of the number of access devices, the pressure of the central network device will increase significantly. In addition, the increase of devices will also cause the expansion of the routing table, so the existing centralized home network is difficult to hold the massive access devices. If there are too many access devices, the network transmission efficiency decreases and the network may be unstable. In a centralized network topology, the central router struggles to handle the access of too many devices. In home networks with a high number of devices, the introduction of new devices leads to increased access latency and communication delays. Moreover, the responsiveness of functions such as service registration and discovery in home networks slows down, and CPU usage escalates.

To solve the problem of decreased communication performance in home networks due to an excessive number of devices, the key lies in adjusting the centralized network topology within home networks. It is necessary to shift from the traditional centralized tree-based networking topology to a distributed mesh networking topology. In complex mesh topology home networks, the IP\cite{postel1981rfc0791} protocol has some inherent issues that are not easily resolved. Firstly, there is the performance overhead issue caused by routing tables. The number of entries in the routing table increases with the network topology, and dynamic routing protocols such as OSPF and BGP are required to maintain the routing table. In home networks, there is a significant disparity in computing power between devices. Both Bluetooth gateways and WiFi routers can function as routers, but there is a large gap in their computing power. Using traditional routing protocols, network performance can be easily affected by devices with lower computing power. Secondly, IP address allocation is quite cumbersome. Each network card involved in IP networking must have a unique IP, and each port on a router needs to be assigned an IP. However, these IPs don't serve many purposes and the allocation of IP addresses is excessively redundant. Using the DHCP\cite{droms1993rfc1541} protocol to allocate an IP can take up to several hundred milliseconds, which increases the latency for device access and roaming. Lastly, IP addresses themselves are related to network topology. When the network topology changes, IP addresses generally need to change accordingly. In home networks, some mobile devices, like smartphones, roam and connect to different wireless routers, meaning the network topology is changing. If the IP protocol is used for networking, the smartphone's IP will change with roaming. This change in IP addresses can cause old TCP connections to disconnect, and the process of obtaining a new IP and re-initiating a TCP connection affects network stability and increases latency.

To address the issues of increased latency and CPU usage caused by the excessive number of devices accessing the network, this paper proposes a new network layer protocol — the Path Vector Header(PVH) Protocol.
The PVH protocol adopts source routing communication, eliminating the need for a routing table to forward data packets. Building on the PVH protocol, we have designed a clustered networking communication system that divides a complex, large network topology into multiple clusters. Each cluster elects a network management cluster. Within each cluster, centralized management is applied to leverage the computational advantages of the network head, while between clusters, distributed management is used to avoid the performance degradation caused by too many centralized network devices. Considering that the IP protocol will remain the mainstream communication protocol for a considerable time to come, and most current applications are based on IP, we have implemented compatibility with the IP protocol through a 'tunneling' method. Compatibility with the IP protocol is achieved using the tun device functionality provided by the Linux kernel. IP addresses are allocated on the tun device, and during communication, IP packets are encapsulated as payload in PVH packets for transmission.
Our core contributions are as follows:
\begin{itemize}
    \item We have designed and implemented a new network layer protocol — the control plane and data plane of the PVH protocol. We have implemented source routing communication, removed routing tables, and achieved table-free forwarding. This solves the performance issues caused by excessive routing table entries and the maintenance of routing tables.
    \item Based on the PVH protocol, we designed and implemented clustering-related algorithms in the PVH control plane. The clustering algorithm can divide a complex mesh network into multiple groups, and within each group, a network head is elected to manage the cluster. Clustering is a way to achieve distributed management, used to address the performance issues of centralized networks.
    \item Building on the clustering of the PVH protocol, we designed and implemented related routing algorithms. Within each group, an improved algorithm based on OSPF is used for routing calculation. The main difference from the original OSPF is that the calculation of the shortest route is carried out by the network head, leveraging the computational advantages of the head and avoiding the involvement of less powerful devices in the calculation of the shortest route. For inter-group communication, a relay broadcast detection method is used, along with caching mechanisms, such as caching path vectors and routing request IDs, to improve inter-group routing performance.
    \item We conducted experimental evaluation tests on the PVH clustering and networking communication algorithm. The experiments show that the PVH protocol can effectively reduce communication latency in the network and also reduce CPU load.
\end{itemize}


\section{Background, motivation and challenge}
\subsection{Background}

\textbf{Smart home.}
With the widespread popularity of smart homes, a large number of intelligent devices have entered ordinary households, bringing a good smart experience to home life. Common smart home devices include terminal devices such as smartphones and tablets, household appliances such as refrigerators and washing machines, and household facilities such as sensors and lighting fixtures.

Smart homes are mainly divided into three parts: smart home systems, smart products, and smart devices. Smart home systems play a role in sensing, linking, and controlling, while smart products mainly endow traditional home products with intelligent functions. Currently, the smart home market is showing a growth trend, and the substantial growth indicates that the smart home industry has entered a turning point, transitioning from a period of hesitation to the next round of integrated evolution. In the next three to five years, smart homes will enter a relatively rapid development stage on the one hand, and on the other hand, protocols and technical standards will begin to actively interoperate and integrate.

The normal operation of smart home systems requires a stable and reliable network environment. Existing centralized home networks overly depend on central nodes, and with the increase in the number of connected devices, the pressure on central nodes will significantly increase. In addition, the increase in devices will also lead to the expansion of routing tables, making existing centralized home networks difficult to handle a massive number of connected devices. When there are too many connected devices, network transmission efficiency will decrease, and the network may experience instability issues.

\textbf{User space networks.}
In some scenarios, there are deficiencies in the relevant code or logic of the kernel network protocol stack. In such cases, it becomes necessary to customize the protocol stack to address these issues, and custom protocol stacks are typically implemented in user space.
A user space network stack is a type of network stack that runs in user space, which means it operates outside of the kernel.

A user space protocol stack is more flexible and secure compared to a kernel space protocol stack. Since the user space protocol stack operates independently of the kernel, it can be updated separately without requiring a full system update when updating the protocol stack. Additionally, user space programs have limited access to kernel resources, making it less prone to causing system crashes or other issues.

Some libraries, such as DPDK\cite{intel2016intel}, netmap\cite{rizzo2012netmap}, provide functionality to decouple the network data path from the kernel. These libraries utilize zero-copy techniques to avoid packet copying between the kernel and user space.

While zero-copy techniques can improve communication efficiency, considering program compatibility and adaptability with the kernel protocol stack, we have retained some packet copying between the kernel and user space.

\textbf{Clustering Network}
Clustering is an important technique in IoT (Internet of Things) networks. In this approach, network nodes are divided into multiple clusters, with each node belonging to and exclusively belonging to a single cluster. Each cluster is overseen by a cluster head, responsible for managing the cluster it belongs to. The primary idea behind clustering is to break down a large and diverse network topology into several smaller, more manageable network systems.
After clustering, communication between devices is categorized into intra-cluster communication (within the same cluster) and inter-cluster communication (between different clusters).

\subsection{Motivation}

\textbf{Home network load optimization.}
Traditional home networks typically employ a centralized networking approach, where the central router in the centralized network experiences excessive load pressure, leading to performance degradation such as increased network latency. Optimizing the home network involves adjusting the centralized tree topology to a distributed network topology, eliminating the presence of a central router. In the optimized structure, devices within the home network can function both as terminal devices for communication and as intermediary devices for packet forwarding.
Some devices in the home network possess significant computational power, such as smart TVs, but this computational power is generally not utilized for networking purposes. To fully leverage the computational resources within the home network, a "clustering" approach can be employed to alleviate the load on the central router. Subsequently, a "selecting cluster head" approach is used to concentrate the computation within the cluster on nodes with stronger computational capabilities. Additionally, adopting source-routing forwarding can avoid the involvement of routing tables during the packet forwarding process, reducing the overhead associated with inter-device packet forwarding.

\textbf{Service-centric Network}
The traditional home network follows a "host-centric" model where the network layer addresses of devices are tied to the network topology. This structure does not align well with the service model in home networks. In a home network, devices such as mobile phones can "roam" between different rooms and routers. The movement of devices results in a change in the network topology, reflected in the IP protocol as a change in subnets. This leads to changes in device IP addresses, requiring a renewal of IP through DHCP, and sockets related to the previous IP need to be recreated, causing fluctuations in network communication latency.
Furthermore, in traditional home networks, every network card used by each device needs to be assigned an IP address. Routers connecting multiple devices through wired connections need to allocate IP addresses to each connected network card, even though these IPs may not be utilized in home network communication. Using too many different IPs to access the same device in a home network can cause confusion and increase the complexity of home network configuration.
In a service-centric network, device network layer addresses are directly associated with the services they provide. Even if device locations change, as long as the provided services remain constant, the device's network layer address remains unchanged. This network structure helps avoid the overhead of DHCP IP renewal and socket recreation, thereby enhancing network stability.

\subsection{Challenge}

\textbf{Clustering networking and communication}
The algorithm aims to divide devices in the network into multiple distinct clusters, with centralized management for intra-cluster devices and distributed management for inter-cluster devices. The execution time of the clustering algorithm should be short, and the hop count between devices within a cluster should not be excessively large. After clustering, communication between devices is categorized into intra-cluster communication and inter-cluster communication, with a priority on ensuring the efficiency of intra-cluster communication.
We employ a completely distributed algorithm for clustering. After exchanging information with other devices within a two-hop range, each device recommends a cluster head based on its capability value. The cluster head then pulls other devices into the cluster. While this clustering method may not always produce the optimal clusters, it operates swiftly, and the resulting clusters meet the criteria of minimizing the hop count between nodes within a cluster.
Following clustering, intra-cluster communication utilizes a variant of OSPF\cite{john1997ospf}. After collecting neighbor information, this information is uploaded to the cluster head, which calculates the routes for communication between nodes based on the cluster's topology. Inter-cluster communication utilizes a broadcast detection method. When communicating with a target node in another cluster, the communication request packet is broadcast to every cluster. The target node, upon receiving the request, responds in a "return path" manner to fulfill the communication request.

\textbf{Source routing forwarding without routing table}
There are various implementations for source routing communication. However, these implementations typically involve the use of routing tables. In home networks, some devices, such as network cameras, may have limited computational and storage capabilities, making it unsuitable to store an excessive number of routing table entries. The existence of routing tables requires a lookup for each packet forwarding, leading to memory query overhead. Additionally, the presence of routing tables complicates the logic of forwarding operations, making it challenging to separate forwarding operations from the protocol stack, for instance, in terms of delegating forwarding operations to network cards.
We adopt a method where the network card identifier is encoded within the source routing, eliminating the need for routing tables. When necessary, such as in Wi-Fi wireless communication and point-to-multipoint wired communication, information about the destination MAC address is encoded in the packet header. This approach involves sacrificing a portion of the packet header space to improve forwarding efficiency.

\textbf{compatibility}
System compatibility is divided into protocol stack compatibility, existing application compatibility, and compatibility across multiple platforms and architectures. Protocol stack compatibility refers to the requirement that the new communication protocol should be mutually compatible with the IP protocol and should not impact existing IP protocols. Existing application compatibility means that the new protocol needs to be compatible with programs previously written based on the TCP/IP protocol stack. Existing programs should be able to use the new protocol for communication without modifying their source code. Compatibility across multiple platforms and architectures mainly involves ensuring that the new protocol can run not only on x86 devices like laptops and desktops but also on ARM devices such as smartphones, routers, and ARM development boards.
We use two new Ethernet packet header protocol numbers (EtherType) to define the data plane and control plane packets for the PVH protocol. Therefore, the new PVH protocol is considered alongside protocols like IPv4 and IPv6\cite{deering2017rfc}, with no mutual interference. We utilize the tun device functionality provided by the Linux kernel to create a virtual network card, ensuring compatibility with existing programs. The packets from programs communicating via the virtual network card are passed to the PVH program, which encapsulates the received IP packets into the payload of PVH data packets using a "tunneling" approach.
Since the PVH protocol-related code is implemented in Go language at the user level, Go's strong cross-platform capabilities make achieving compatibility across multiple platforms and architectures relatively straightforward. For x86 devices, installing necessary libraries is sufficient without requiring many additional operations. For routers, ARM development boards, etc., it is necessary to compile OpenWRT-related code to ensure environmental consistency. Root permissions are needed for smartphones to run the PVH protocol, and manual compilation and installation of some dynamic link libraries may be required.

\section{Overview}
\begin{figure}[htbp]
    \centerline{\includegraphics[scale=0.8]{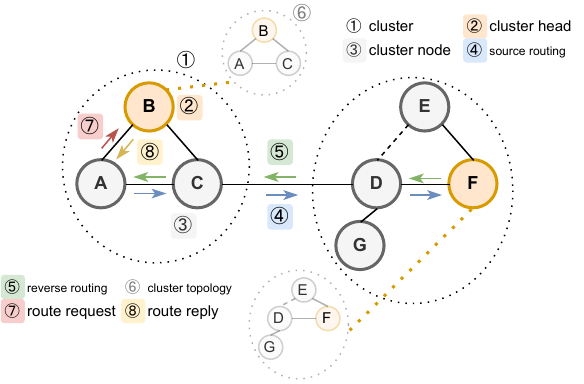}}
    \caption{PVH overview}
    \label{pvh_overview}
\end{figure}

The packet networking system is mainly divided into a control plane and a data plane. The control plane data packets focus on programmability, and use protobuf to encode the content of some data packets. The data plane focuses on efficiency, and the data packets adopt the direct encoding method.
The control plane is mainly responsible for the establishment of the cluster network, route calculation, cluster topology maintenance, node online and offline, and reverse routing construction. The data plane is mainly responsible for the transmission of IP compatible data packets in tunnel mode.
The control plane and the data plane belong to the same process, and the communication between the two uses the built-in channel or shared variable of the go language.

The cluster network algorithm is a distributed parallel algorithm. When the cluster is initialized, all nodes will participate in the calculation of the cluster initialization. cluster maintenance is a process of updating the cache regularly, each node maintains its own neighbor table and uploads the network head periodically. When a node goes online, it can actively scan nearby clusters and join them. When a node goes offline, the cache of the node in the network head expires.
The forwarding of the data packet is communicated by means of source routing. The header encodes a path vector, which stores the number of the network card that the node needs to pass through. During communication, the path vector back to the source node can be obtained by constructing a reverse route based on the first data packet. The data packet is transmitted by encapsulating the IP data packet into the payload of the PVH data packet. After the data packet reaches the destination node, the PVH packet header is removed, and the internal IP data packet is injected into the kernel.

\section{Detail}
\subsection{Data Plane}
This project has designed a unique PVH (Path Vector Header) network protocol to meet the requirements of the data plane, and has achieved efficient and convenient data plane transmission functionality through the unique "path vector forwarding" routing mechanism. In order to be compatible with existing network protocols, this project has also implemented tunneling communication technology for IP protocol on top of the PVH network protocol.

The path vector header (PVH) is a new type of network layer header used in the data plane, which can be used independently for data transmission using the PVH protocol, or can be used in "PVH tunnel mode" to be compatible with IP protocol communication. In the "PVH tunnel mode", IP data packets are transmitted as payload within PVH data packets.

\begin{figure}[htbp]
    \centerline{\includegraphics[]{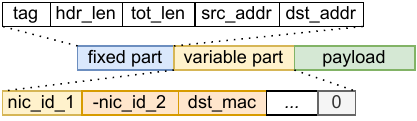}}
    \caption{PVH structure}
    \label{pvh_packet}
\end{figure}

As shown in figure \ref{pvh_packet}, the PVH packet header occupies at least 17 bytes. 
The PVH field encodes path vectors, which are recorded in the packet header in the form of source routing. Each element in the path vector represents the next-hop port from which the packet should be forwarded. For detailed forwarding rules, please refer to section \ref{sec:forwarding_mechanism}. The path vector in the header is encoded using one byte per hop, and in practical clustering scenarios, the number of hops between two nodes typically does not exceed 8 hops. Therefore, the PVH data packet header generally does not exceed 25 bytes. The "tag" field in the header indicates the category of the data packet. For example, a tag of 0x01 indicates that the data packet is operating in PVH tunnel mode, and the payload is a valid IP data packet. The "hdr\_len" field indicates the length of the data packet header, while the "tot\_len" field indicates the total length of the data packet. The "src\_addr" and "dst\_addr" are both 6-byte network layer addresses, which serve as unique identifiers for a host. "src\_addr" represents the source host address of the data packet, and "dst\_addr" represents the destination host address of the data packet. The path vector contains the information needed for data packet forwarding.

\subsection{Forwarding Mechanism}\label{sec:forwarding_mechanism}

In cluster-based self-organizing networks, nodes do not have routing tables, and each node can act as both a "terminal node" to provide services or initiate service requests, and as a "router" to forward packets. Data communication between nodes is based on source routing, where the path vector for forwarding is calculated and included in the PVH packet header before sending. Intermediate nodes then forward the packet based on the path vector in the packet.

During packet forwarding, there are two types of links involved, namely point-to-point links and shared media links. For example, when two network cards are directly connected by an Ethernet cable, it forms a point-to-point link. When multiple network cards are connected through a switch, or wireless network cards are connected together through an access point (AP), it forms a shared media link.

\begin{figure}[htbp]
    \centerline{\includegraphics[]{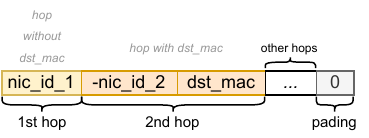}}
    \caption{Path Vector}
    \label{path_vector}
\end{figure}

As shown in figure \ref{path_vector}, the node forwards the packet based on the data in the path vector of the packet, distinguishing between point-to-point link and shared medium link during forwarding.

When encoding the path vector, the point-to-point link is represented as a positive 8-bit signed integer in the path vector. This integer indicates from which numbered network card the packet should be sent out. When the packet is sent out, the destination MAC address at the data link layer is set to ff:ff:ff:ff:ff:ff, which is the broadcast MAC address. Since it is a point-to-point link, when the packet is broadcasted to the next hop, at most one device will receive the broadcasted packet.

On the other hand, the shared medium link is represented as a negative 8-bit signed integer in the path vector, where the negation of the integer indicates from which numbered network card the packet should be sent out. In addition, the path vector header for shared medium link includes a destination MAC address, which is the MAC address of the data link layer when the packet is sent out. With the destination MAC address provided, when the packet is sent to the next hop, at most one device will receive the packet.

\begin{figure}[htbp]
    \centerline{\includegraphics[]{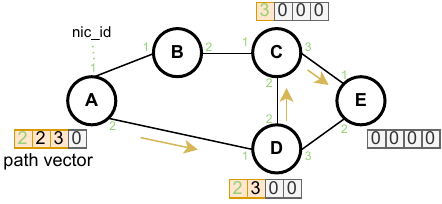}}
    \caption{Packet Forwarding}
    \label{packet_forwarding}
\end{figure}

As shown in figure \ref{packet_forwarding}, when Node A communicates with Node E, before sending the first packet, routing calculation is performed within the cluster or between clusters to obtain the path vector from A to E: [2, 2, 3, 0]. The leading element of the path vector is 2, indicating that the next hop for the packet should be Node A's interface 2. When the packet is sent, the path vector is shifted left as a whole, and the leading element is popped out, so when Node D receives the packet, the path vector becomes [2, 3, 0, 0]. Similarly, the packet is sent from Node D's interface 2 to reach Node C, and then from Node C's interface 3 to reach Node E. When the packet arrives at Node E, the destination address matches E's address, indicating that the packet has reached the destination. Node E receives and processes the packet.

\begin{algorithm}
\caption{Packet forwarding}
$pv \gets \text{path vector}$
$n \gets \text{pv[0]}$\;
\While{true}{
    \eIf{$n == 0$} 
    {
        \text{packet arrived at destination host}\;
    }
    {
        \text{left shift path vector by one byte}\;
        \eIf{$n > 0$}{
            \text{
                write\_packet$(NIC_ID=n, dmac=broadcast)$\;
            }
        }{
            $dmac \gets \text{pv[:6]}$
            \text{left shift path vector by 6 bytes}\;
            \text{
                write\_packet$(NIC\_ID=n, dmac=dmac)$\;
            }
        }
    }
}
\end{algorithm}

\subsection{Compatible with IP Protocol}
The system operates on a principle similar to Virtual Private Network(VPN) and is compatible with the IP protocol. To achieve the functionality of encapsulating one packet within the payload of another, packets are sent out from a "tun" device. The "tun" device functions as a virtual network interface, and packets transmitted from the "tun" device are directed to a user space program, which, in this case, is the PVH program. The PVH program encapsulates the IP packets received from the "tun" device into PVH packets and transmits them through a regular network interface to other nodes in the network. Nodes that provide services based on the IP protocol, such as servers offering HTTP services, and nodes initiating requests for IP services, like accessing a website homepage, need to configure IP addresses. Other nodes do not require IP addresses.

To enable compatibility with IP communication, some preparatory steps need to be taken. You can use the "ip tuntap add" command to create a "tun" device. In the PVH program configuration, specify the name of the corresponding "tun" device. Then, assign an IP address to the "tun" device, with a subnet mask length of 32 to avoid the generation of ARP packets during communication. Finally, you can modify the routing using the "ip route" command to ensure that packets destined for specific network segments are transmitted via the "tun" device.

\begin{figure}[htbp]
    \centerline{\includegraphics[]{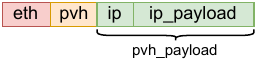}}
    \caption{IP-over-PVH Packet}
    \label{ip_over_pvh}
\end{figure}

During communication, packets sent from the "tun" device are handed over to the PVH program for processing. The PVH program will encapsulate the outer layer of the IP packet with a PVH header, creating an IP-over-PVH packet. These IP-over-PVH packets are forwarded throughout the network using PVH source routing. When the packets reach their destination host, the PVH program on the destination host extracts the payload of the IP-over-PVH packet, which is an IP data packet. This IP data packet is then delivered to the kernel protocol stack through the "tun" device for further processing by the kernel.

\subsection{Control Plane}

The emphasis on the control plane and data plane differs in their respective priorities. For the data plane, communication efficiency, i.e., the speed of processing packets, is crucial. On the other hand, for the control plane, programmability takes precedence. As a trade-off, some performance can be sacrificed in the control plane. This is why control plane packets use the Protocol Buffers (protobuf) library for encoding and decoding control information.

\begin{figure}[htbp]
    \centerline{\includegraphics[]{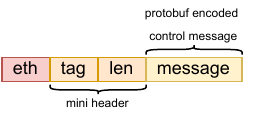}}
    \caption{Control Plane Packet}
    \label{control_plane_packet}
\end{figure}

Due to the minimum frame size limitation of Ethernet, any shortfall in packet size will be automatically padded with zeros to reach a minimum of 64 bytes. However, control plane information may not necessarily end without zeros, which makes it unsuitable to directly use Ethernet protocol for sending control plane data, as zero padding may render the information invalid. Therefore, as shown in figure \ref{control_plane_packet}, we have designed an additional 4-byte field in the Ethernet packet header, where the first two bits indicate the decoding method and the last two bits indicate the length of the packet. When facing automatic zero padding, we can use this length field to extract the valid information, thus resolving the zero padding issue in the native Ethernet protocol.

\subsection{Cluster Initialization}
The objective of the clustering establishment algorithm is to quickly form sub-clusters given a set of nodes. Since this algorithm prioritizes speed of establishment over the optimality of "maximizing intra-cluster traffic and minimizing inter-cluster traffic", the formed sub-clusters may not be optimal. If necessary, the sub-clusters can be modified based on the algorithms provided in the next section on cluster maintenance.

In order to elect the head node during the process of cluster establishment, we have designed a metric called "capability value".

$$
N = \alpha C + \beta M + \gamma B, \quad \alpha + \beta + \gamma = 1
$$

The "capability value" is a metric that takes into account various factors such as node processing power (C), memory size (M), and total bandwidth of the network interfaces participating in the clustering (B), denoted as N. The capability value is used to assess the overall processing power of a node, where higher processing power indicates that a node is more suitable to be elected as the head node in the cluster formation process.

The clustering formation process mainly consists of three stages: the broadcast stage, the head node election stage, and the clustering scanning stage.

During the broadcast stage, each node broadcasts its own capability value to other nodes through multi-hop communication. Upon receiving the capability values of other nodes, each node stores them along with the MAC addresses of the respective nodes for subsequent calculations. At the end of this stage, each node will have obtained the capability values of all the nodes within a certain range (X hops) centered around itself.

During the network head (net-head) election stage, each node queries the capability value information of other nodes within the X hop range that it has stored. If a node has one of the highest capability values among the nodes within the X hop range, it becomes the net-head. The net-head then broadcasts a net-head declaration to the nodes within the X hop range.

During the clustering scanning stage, for nodes that did not join any cluster in the previous stage, they need to send cluster scan packets to their neighbors. When a neighbor receives a cluster scan packet, it replies with information about the cluter it belongs to. The node that did not join any cluster selects one of the scanned clusters and joins it based on the received information.

\begin{figure}[htbp]
    \centerline{\includegraphics[]{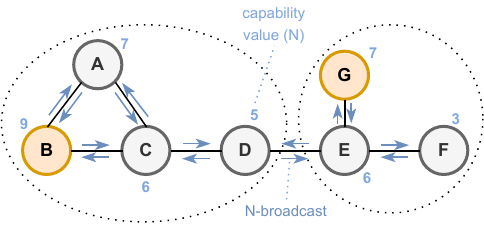}}
    \caption{Cluster after init}
    \label{cluster_init}
\end{figure}

As shown in Figure \ref{cluster_init}, during the cluster initialization phase, nodes initially broadcast their capability values to nodes within a hop range of X=2. After a certain period, the node with the highest capability value becomes the network leader. The network leader sends out invitation packets to invite other nodes to join the cluster. Nodes that do not receive invitations can also proactively join the cluster. After cluster initialization, the nodes are divided into two different clusters.

\subsection{Cluster Maintenance}
The nodes in a cluster send neighbor broadcasts to their neighbors at regular intervals. When a node receives a neighbor broadcast, it records the neighbor's information in a neighbor table, including the neighbor's address and the network port through which they are connected. If the connection is through a shared link port, the source MAC address in the Ethernet layer of the neighbor's broadcast packet is also recorded.

\begin{figure}[htbp]
    \centerline{\includegraphics[]{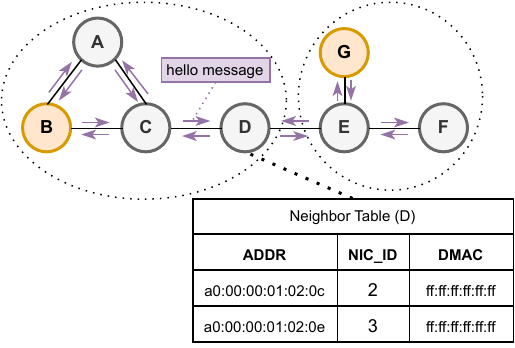}}
    \caption{Neighbor Broadcasting}
    \label{neighbor_broadcast}
\end{figure}

As shown in Figure \ref{neighbor_broadcast}, each node in the network periodically sends hello packets\cite{john1997ospf} to its neighboring nodes. Nodes that receive these hello packets add the neighbor's information to their neighbor table. This neighbor information includes the neighbor's network layer address (ADDR), a 48-bit address assigned by the PVH program, the NIC(Network Interface Card) ID of the receiving network card, and, if the connection is not point-to-point, the source MAC address is recorded in the DMAC field. The meaning of a neighbor table entry is that "if a packet is sent from the network card with ID NIC\_ID, and the destination MAC address is set to DMAC (for entries that do not contain DMAC, the destination MAC address is the broadcast address), the packet will reach the node with the address ADDR."

After each neighbor broadcast, a random wait time of 1 to 2 second is introduced. After the wait time, the node uploads its collected neighbor table to the cluster head. The neighbor tables received by the cluster head can be treated as an "adjacency list" in graph theory. The cluster head can then reconstruct the topology of the cluster based on the received neighbor tables.

\begin{figure}[htbp]
    \centerline{\includegraphics[]{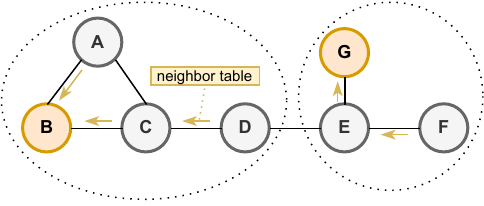}}
    \caption{Upload Neighbor Table}
    \label{upload_neighbor_table}
\end{figure}

As shown in Figure \ref{upload_neighbor_table}, each node periodically uploads its neighbor table to the network leader. The network leader can collect neighbor tables from all nodes in the entire cluster. The neighbor tables contain information about the "edges" in the entire network topology. Based on this edge information, the cluster topology can be established. The network leader uses the cluster topology information it has to compute intra-group routing.

When a new node joins the cluster networking system, it first uses the cluster scanning to identify the existing clusters. The new node can then choose to join one of the scanned clusters. Upon receiving a cluster scanning data packet, if the node has already joined a cluster, it constructs a reverse route and replies with the cluster head information to the initiator of the cluster scanning. If the node has not joined any cluster, it relays the cluster scanning data packet. Once the new node joins a cluster, it starts periodic neighbor broadcasting and uploads its neighbor table. When the cluster head receives the new neighbor table, it uses it to reconstruct the cluster topology, considering the newly joined node in the topology.

When a node goes offline, it stops sending neighbor table update packets to the cluster head. If the cluster head does not receive neighbor table updates from a node for a certain period of time, it considers the node as offline. The adjacency table entries in the sub-cluster topology that are associated with the offline node will be marked as expired, and the routing calculations within the cluster will no longer consider the offline node. This ensures the accuracy of the sub-cluster topology information, so that other nodes can perform routing calculations and communications without being affected by the offline node.

\subsection{Routing Calculation}

\subsubsection{Reverse Routing}

If node X sends a data packet to node Y using a certain form of communication such as relay broadcast or PVH path vector forwarding, node Y expects to obtain a reverse route from Y to X through a certain calculation. The purpose of constructing a reverse route is to obtain this reverse routing information.

Each time a data packet is forwarded, the source of the data packet is recorded to calculate the reverse route. For point-to-point links, the recorded information is the outgoing network interface index, which indicates which port the packet was forwarded through. For shared media links, the source MAC address of the packet is also recorded. When a data packet with reverse routing information is sent, each intermediate node along the path adds its corresponding port to the path recorded in the data packet. When the receiving node receives the packet, it simply reverses the path to obtain the reverse route.

\begin{figure}[htbp]
    \centerline{\includegraphics[]{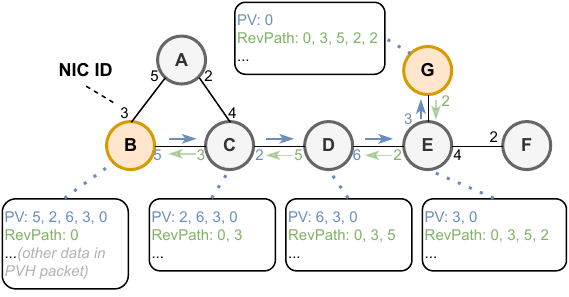}}
    \caption{Reverse Routing}
    \label{reverse_routing}
\end{figure}

In Figure \ref{reverse_routing}, the numbers next to the nodes represent the NIC (Network Interface Card) IDs. 
Assuming that node B already knows the route to node G in advance, when Node B communicates with Node G, the first data packet is used to establish routing. The "Pv" field in the data packet indicates the forward route, which is the source route used for packet forwarding. The "RevPath" field in the data packet represents the reverse route, which is used to construct the route for the packet to return to the sender. During packet forwarding, when each node receives the data packet, it appends the NIC ID of the receiving NIC to the end of the "RevPath" and then reverses the "RevPath." This process allows you to obtain the route from the current node back to the sender. Node G, after receiving the initial data packet sent by Node B, reverses the "RevPath" to obtain the route from G to B, which is 2, 2, 5, 3, 0. This route represents the source route "PV" from G to B.

\subsubsection{Routing in Cluster}
All intra-cluster routing is calculated by the cluster head of the respective sub-cluster. As the cluster head has the entire topology information of the sub-cluster, if a node within the cluster needs to compute an intra-cluster route, it sends a routing request to the cluster head. Upon receiving the routing request, the cluster head uses a breadth-first search (BFS) algorithm to calculate the shortest path (assuming all edge weights in the sub-cluster topology graph are 1). In this case, BFS is more efficient than other algorithms such as SPFA or Dijkstra. Once the calculation is completed, the cluster head sends the routing result to the node that requested it.

\subsubsection{Routing between Clusters}
In general, communication and traffic related to routing queries are limited within the cluster. However, in some cases, cross-cluster communication is required. During cross-cluster communication, the cluster boundaries are temporarily ignored, and a relayed broadcast probing method is used to find the target node. Once the target node is found, a reverse route is constructed and the result is returned. During the return of the routing, the reverse route information is recorded. When the routing response data packet reaches the source node, the reverse route is constructed, resulting in the route from the source node to the target node. To avoid excessive relayed broadcast data packets in the network, the routing request data packets contain an ID for deduplication. The "key" of the data packet, which is composed of the ID and the source address of the routing request, is used for deduplication. Data packets with the same "key" are considered as the same cross-cluster routing request. For each node, at most one relayed broadcast of the same cross-cluster routing request is allowed within a certain period of time.

\begin{figure}[htbp]
    \centerline{\includegraphics[]{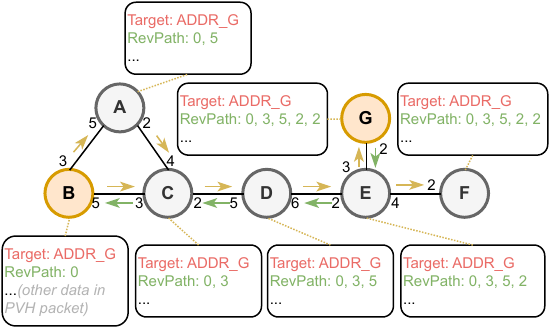}}
    \caption{Relay Broadcasting}
    \label{relay_broadcasting}
\end{figure}

As depicted in Figure \ref{relay_broadcasting}, assuming that node B and node G are not in the same group and need to communicate, node B can send a relay broadcast packet to each of its neighbors with a "TargetAddr" set to the address of node G. Every node in the network that receives this relay broadcast packet will rebroadcast it. The data packet is tagged with an ID to ensure that each node processes it at most once.

After the relay broadcast, every node in the network will receive this data packet. When the destination node G receives the data packet, it reverses the "RevPath" to obtain the route from G to B and sends a response packet to node B. The response packet also constructs a "RevPath" as it travels from G to B. Upon receiving the response, node B reverses the "RevPath" to obtain the route from B to G.

\section{Evaluation}
The evaluation of the project primarily aims to demonstrate and validate the functionality and performance of the system. On one hand, to assess the availability and usability of the system, tests are conducted using physical routers, development boards, and other hardware components. On the other hand, in order to achieve better testing results and showcase the internal operations of the system, virtual environments are used to simulate multi-node communication for testing purposes.

\begin{figure}[htbp]
    \centerline{\includegraphics[]{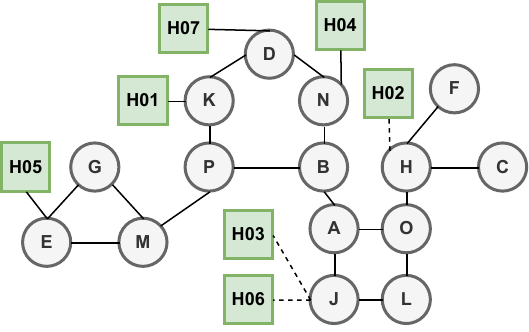}}
    \caption{Physical Environment Topology}
    \label{physical_topo}
\end{figure}

In our built hardware environment, as shown in figure \ref{physical_topo}, there are seven terminal nodes consisting of three laptops, one desktop computer, one mobile phone, and two development boards running Android OS. The testing environment also includes 15 routers, which are organized into a topology resembling four independent sub-clusters. The connections between router nodes are wired point-to-point links, while the connections between terminal nodes and router nodes are a combination of wired point-to-point links and shared media links (wireless networks).


In addition, we also conducted partial performance testing of the home sub-cluster self-organizing network in an emulated environment. In the emulated environment, the system topology was either randomly generated with varying structure and scale, or simulated to resemble typical large-scale building network topologies.

\subsection{Emulated Environment}
In the emulated environment, we tested the communication delay between nodes, service delay, and CPU usage. Then, on this basis, network topologies of various scales are randomly generated, and the variation of node communication delay with network scale is tested.

First, we test various indicators of a emulated typical home network, such as communication delay, service delay, cpu usage, etc. A typical home network has 19 nodes, and the nodes include wired and wireless routers, desktop computers, Notebooks, smart lights, mobile phones, smart switches, smart TVs, etc. A connected graph is formed between nodes through wired and wireless hybrid networking, and includes at least one ring.

\begin{figure}[htbp]
    \centerline{\includegraphics[scale=0.4]{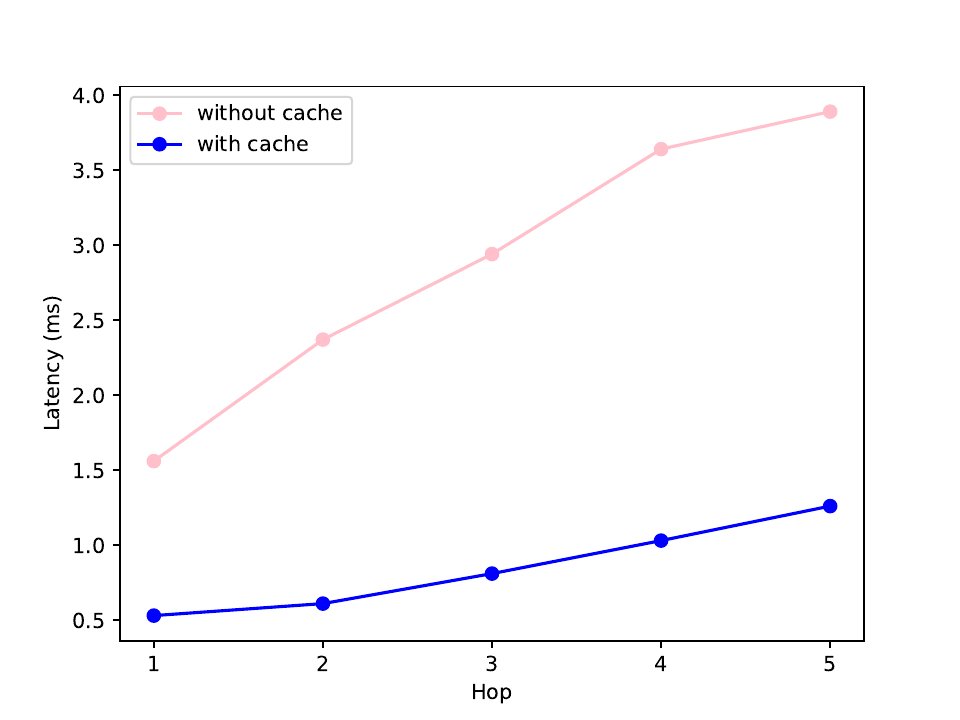}}
    \caption{Ping Test in Emulated Environment}
    \label{emu_ping}
\end{figure}

As shown in figure \ref{emu_ping}, the node communication delay does not change much with the number of hops, and with the increase of the number of hops, the increase of communication delay is not large. When the nodes communicate for the first time, the route needs to be calculated, so the delay will be relatively high. Subsequent communications cache the route and communicate using the cached route.

\begin{figure}[htbp]
    \centerline{\includegraphics[scale=0.4]{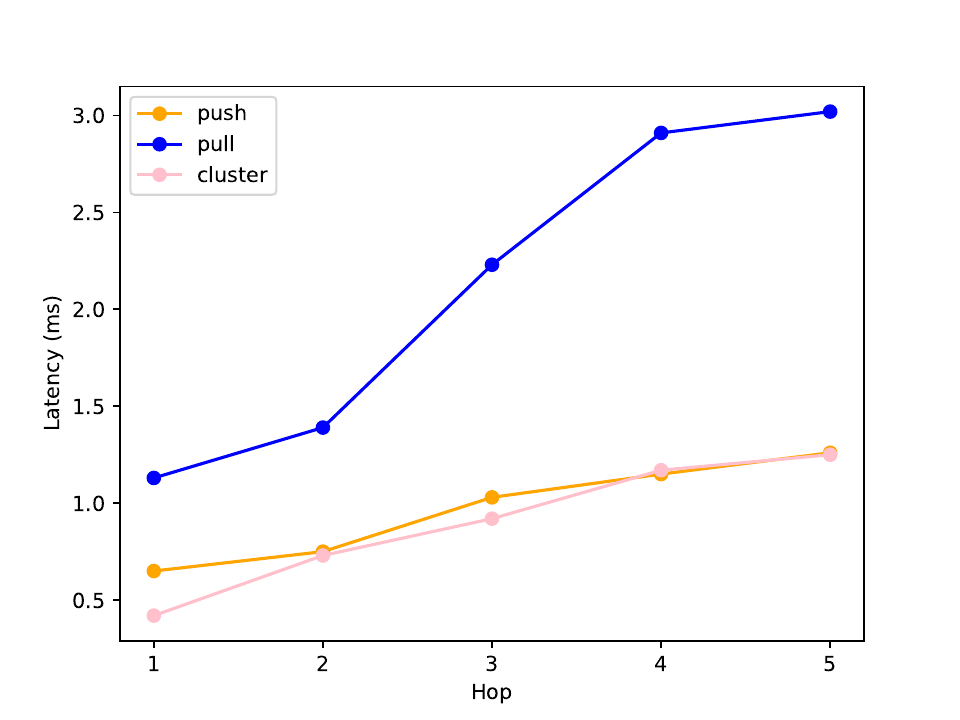}}
    \caption{Service latency Test in Emulated Environment}
    \label{srv_test}
\end{figure}

The push method of service publishing means that the node actively broadcasts and pushes its own service. The pull method means that the node that needs to access the service first initiates a request to the server, and the server replies to the request initiator with service information after receiving the request.

As shown in figure \ref{srv_test}, The service delay of the cluster network is similar to that of the traditional push method, but lower than that of the pull method. With the increase of the number of hops, the delay is relatively stable and the delay does not increase much.

In order to make the difference in CPU usage more obvious, in the CPU usage test, we publish about 1000 services, and randomly select 5 nodes among them to initiate access to the service, and test the CPU usage in the whole process.

\begin{figure}[htbp]
    \centerline{\includegraphics[scale=0.4]{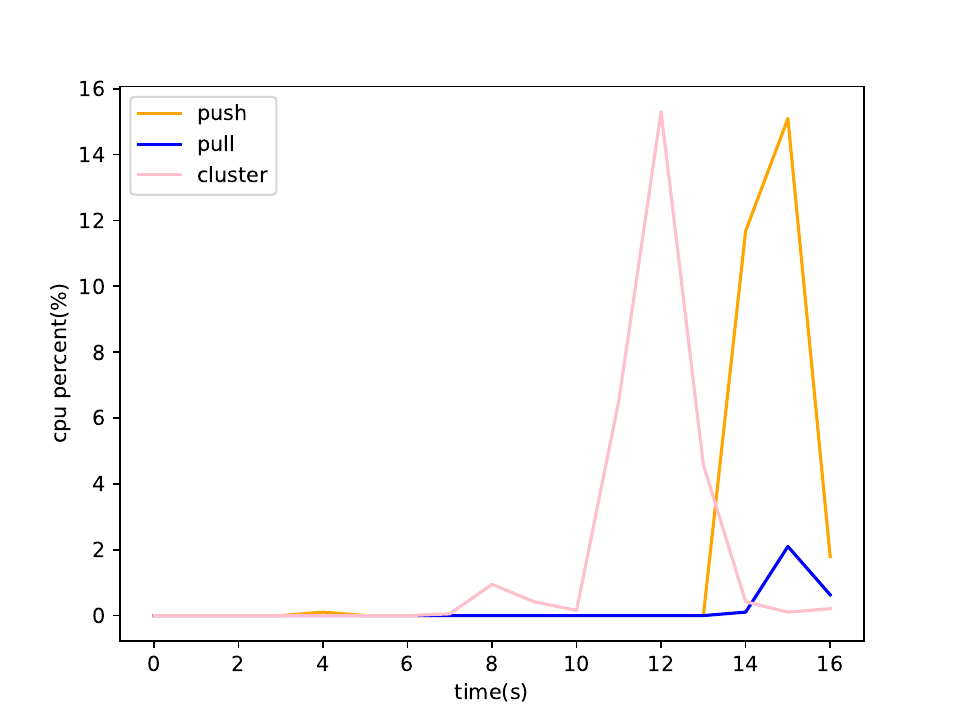}}
    \caption{CPU Usage Test - Stage 1}
    \label{cpu_stage1}
\end{figure}

As shown in figure \ref{cpu_stage1}, the main CPU usage of the cluster network comes from cluster initialization and service registration and push, the main CPU usage of the push method comes from the registration and push of the service, and the main CPU usage of the pull method mainly comes from the registration of the service. Therefore, the CPU usage of the cluster network will have two peaks, the lower peak comes from the cluster network initialization control packet, and the higher peak comes from the registration and push of the service. The CPU usage of the Push method comes from the registration and push of the service, while the Pull method only needs to register the service and does not need to push the service, so it occupies the least CPU resources.

\begin{figure}[htbp]
    \centerline{\includegraphics[scale=0.4]{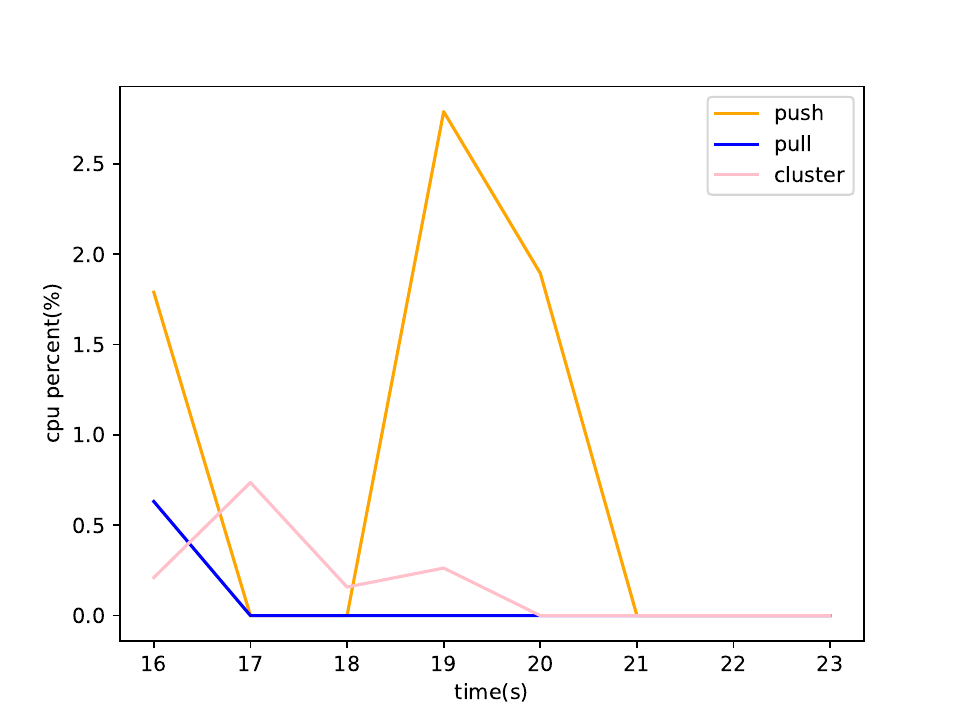}}
    \caption{CPU Usage Test - Stage 2}
    \label{cpu_stage2}
\end{figure}

As shown in figure \ref{cpu_stage2}, at the 2nd stage, the cluster network method needs to maintain the network topology by uploading the neighbor table to the network head, so there is a little CPU energy consumption. The push method creates multiple keep-alive threads at this stage, which consumes the most energy. The method requires no additional operations at this stage and consumes the least amount of energy.

\begin{figure}[htbp]
    \centerline{\includegraphics[scale=0.4]{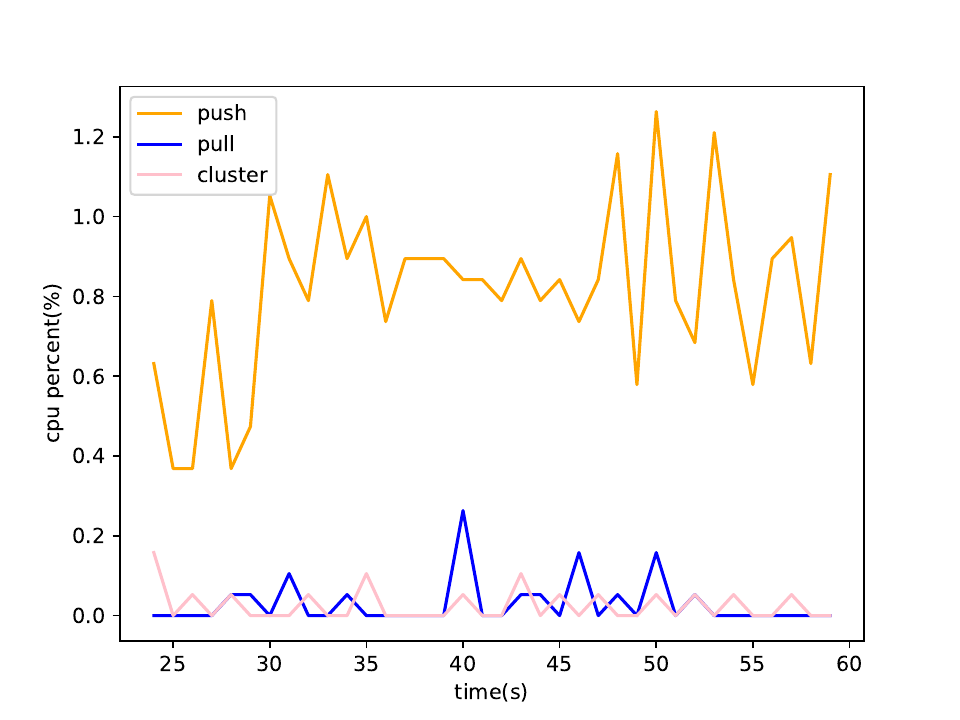}}
    \caption{CPU Usage Test - Stage 3}
    \label{cpu_stage3}
\end{figure}

As shown in figure \ref{cpu_stage3}, at the 3rd stage, randomly select 5 nodes to initiate service access. The energy consumption of the cluster network algorithm at this stage comes from the cluster network control data packet, which has the lowest energy consumption. The push method needs to be kept alive regularly, which consumes a lot of energy. The pull method is When serving queries and replies, it is necessary to consider the network as a whole to calculate routing, which also consumes a certain amount of energy.

\subsection{Physical Environment}

The test environment is built in kind, which includes development boards, routers, notebooks, desktops, mobile phones, etc., and a wired and wireless mixed network is used to form a network topology with rings. Repeat the following steps several times: determine a number of hops X, randomly select two nodes, try to ensure that the number of hops between the two nodes is X, and then conduct multiple communication delay tests.

\begin{figure}[htbp]
    \centerline{\includegraphics[scale=0.4]{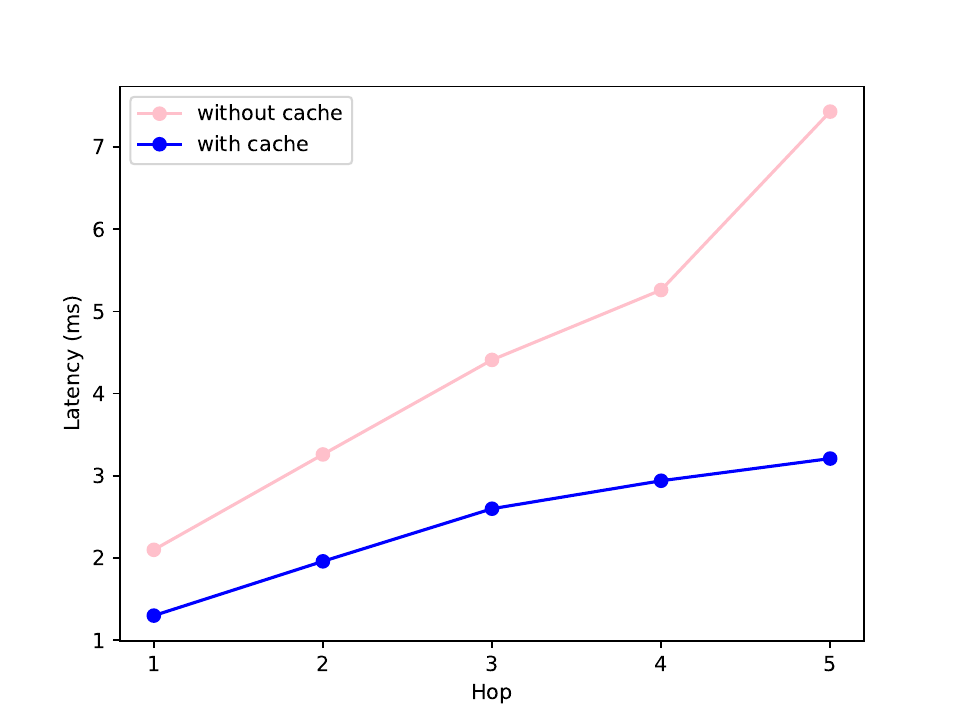}}
    \caption{Physical Latency without Wireless Hop}
    \label{latency_without_wireless}
\end{figure}

As shown inf figure \ref{latency_without_wireless}, the communication delay between nodes is generally relatively low (less than 10ms). When multiple communication tests are performed between the same two nodes, only the first communication delay is relatively high (need to calculate the route), and the subsequent communication will use the cached route. Communication, before the cache expires, the communication delay will be relatively low.

\begin{figure}[htbp]
    \centerline{\includegraphics[scale=0.4]{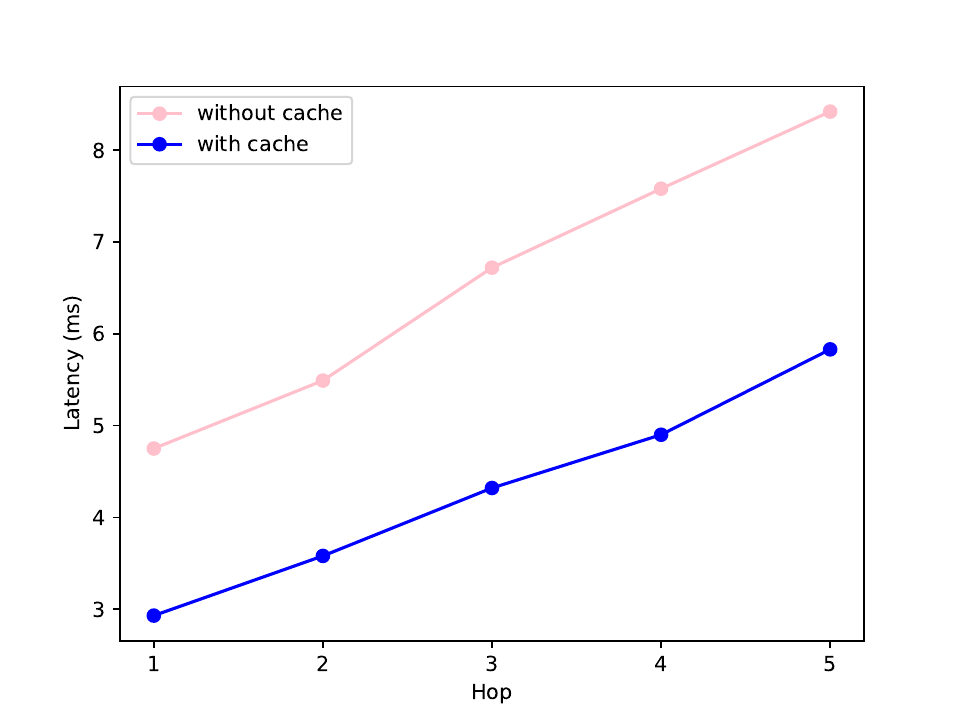}}
    \caption{Physical Latency with Wireless Hop}
    \label{latency_with_wireless}
\end{figure}

Introducing one wireless hop in the communication delay test in the physical environment, as shown in figure \ref{latency_with_wireless}, although the communication delay will increase to a certain extent, it will not have much impact.

\section{Related Work}

\textbf{Network architecture and communication methods}
Research on network architecture and communication methods in networks is abundant, but much of this research is not directly applicable to home networks. There are two main reasons for this. Firstly, there are compatibility issues. Many protocols\cite{clark2003new, zhang2011scion, andersen2008accountable, seskar2011mobilityfirst, nordstrom2012serval, godfrey2009pathlet, yang2007nira}, while improving network communication efficiency, are incompatible with the TCP/IP protocol. Some older applications may require source code modifications to use new protocols.mTCP\cite{jeong2014mtcp} and mOS\cite{jamshed2017mos} employ the run-to-completion model to tightly couple programs with the protocol stack. IX\cite{belay2014ix} modifies the system's network API semantics. Moreover, some software is closed-source, making it challenging to modify the source code.
Secondly, there are challenges in dealing with the IP protocol. Due to the narrow waist structure of the Internet, most adjustments to network structure and protocols occur at the network layer. Some protocols add a 3.5 layer above the IP layer to extend the functionality of the network layer\cite{mccauley2019enabling, nordstrom2012serval}. While this approach is convenient, it retains some mechanisms and deficiencies of the IP layer, such as the need to change IP addresses with changes in network topology and the reliance on routing tables for packet forwarding.
In the context of a home network environment, a more suitable approach would be to add a 2.5 layer below the IP layer, creating an overlay on the IP layer. The IP layer exists for compatibility with existing applications based on the TCP/IP protocol stack, but actual communication occurs using the 2.5 layer. From another perspective, the 2.5 layer protocol can be seen as a new network layer protocol, with IP packets serving as the payload for this new protocol. In other words, IP packets are encapsulated in the new protocol's packets using a tunneling approach\cite{gilligan2005basic, conta1998generic}.

Some user-space protocols, such as Light\cite{li2023light}, utilize functions with the same signatures as system calls and modify LD\_PRELOAD to achieve transparent compatibility with applications. However, a drawback of this compatibility approach is that the relevant programs need to dynamically link the glibc library during compilation, making it incompatible with statically linked programs.
In contrast, the PVH protocol creates a virtual network interface using the tun device provided by the kernel. It assigns an IP to the virtual interface to achieve compatibility with existing TCP/IP programs. This approach supports compatibility with almost all programs, without the limitation of being incompatible with statically linked programs. The downside of using the tun device is that there are more system calls during communication, involving multiple copies of data packets, which can result in relatively lower performance.

\textbf{Network clustering}
Adopting a clustering approach to divide devices in the network into multiple distinct clusters, each electing a cluster head to oversee and manage the cluster collectively. This clustering strategy is widely applied in the communication of many wireless self-organizing networks, with various clustering and communication algorithms associated with it\cite{AMGOTH2015357, jaiswal2020eomr, 926982}. However, there are significant differences between home networks and general IoT networks or wireless self-organizing networks. These differences primarily stem from two aspects: first, devices in a home network are interconnected through a mix of wired and wireless connections, with the core of the home network being wired. Issues related to limited battery capacity or power causing a decline\cite{badi2021reapiot} in communication performance are virtually nonexistent. Second, devices in a home network exhibit heterogeneity, with substantial differences between them. For instance, the computational power of a smart TV surpasses that of a network camera by a significant margin. Therefore, during clustering and communication, it is necessary to assign more tasks to cluster heads with higher computational power.

\section{Conclusion}
This project proposes a self-developed architecture for home sub-cluster self-organizing networks, which is based on sub-cluster self-organizing networks and maintains good performance and scalability even under conditions of massive device access, addressing the pain points of current home network environments struggling to support large-scale device connections. The project also builds a test environment that can simulate large-scale home networks and implements a prototype system in this environment to demonstrate its performance. We have developed a new communication protocol for home networks called the PVH protocol. This new protocol addresses the issue of poor network communication performance when dealing with complex mesh network topologies. On one hand, the PVH protocol employs 48-bit addresses instead of 32-bit addresses, allowing devices to use their MAC addresses directly as their network addresses. This eliminates the need for address allocation and, consequently, the overhead associated with DHCP.On the other hand, the new protocol eliminates the concept of subnets and routing tables, using source routing for communication. This solves the problem of having too many devices in a network, which could result in an excessive number of routing table entries. Tableless forwarding provides high-performance packet forwarding. Furthermore, the protocol can seamlessly integrate with the IP protocol through tunneling, offering transparent compatibility with higher-level applications. By using tunneling technology, applications based on the IP protocol can communicate using the PVH protocol without needing any modifications.

Firstly, in the context of home environments and massive device access, this project has specifically designed a data plane based on the self-developed "pvh" network protocol, and implemented a home sub-cluster self-organizing network on top of it. Meanwhile, the communication protocol for the control plane has been designed in a reasonable manner to facilitate network configuration.

Furthermore, in accordance with the project's objectives and requirements, we have specifically designed unique mechanisms for cluster node online/offline, topology establishment and maintenance, routing calculation, and path vector forwarding. These mechanisms have enabled the system to achieve the predefined performance indicators.

Finally, we have successfully completed a prototype system for this project. With the support of the prototype system, we were able to achieve massive device access and usage. To validate the performance of our system, we also built a hardware testing platform and a virtual simulation environment, where we conducted tests and demonstrated the network's capability for massive device access and excellent communication performance.

\section{Future Work}

Currently, the code implementation is user-space-based, which involves multiple system calls for packet processing and lacks support for packet fragmentation and reassembly in the kernel or network card. Although the protocol stack has low latency, its throughput is also relatively low. Future work involves moving the protocol stack into the kernel to reduce the overhead of system calls and data packet copying.

Additionally, since data packets are forwarded throughout the entire network using source routing without the need for table lookups, the packet forwarding logic can be offloaded to programmable network cards. Offloading the forwarding logic can reduce CPU overhead and save memory resources.

\printbibliography

\end{document}